\documentclass[aps,12pt,final,notitlepage,oneside,onecolumn,nobibnotes,nofootinbib,superscriptaddress,showpacs,centertags]{revtex4}
\usepackage{amssymb,latexsym,amsmath,amsbsy}
\usepackage{young}
\DeclareMathOperator{\ch}{char}
\DeclareMathOperator{\sdim}{sdim}

\begin{document}

\title{On characters and superdimensions of some infinite-dimensional\\
irreducible representations of $\mathfrak{osp}(m|n)$}
\author{\firstname{N.I.}~\surname{Stoilova}}
\affiliation{Institute for Nuclear Research and Nuclear Energy,
Boul.\ Tsarigradsko Chaussee 72, 1784 Sofia, Bulgaria}
\author{\firstname{J.}~\surname{Thierry-Mieg}}
\affiliation{NCBI, National Library of Medicine, National Institute of Health,
8600 Rockville Pike, Bethesda MD20894, USA}
\author{\firstname{J.}~\surname{Van der Jeugt}}
\affiliation{Department of Applied Mathematics, Computer Science and Statistics, Ghent University, Belgium}

\begin{abstract}
Chiral spinors and self dual tensors of the Lie superalgebra $\mathfrak{osp}(m|n)$ are infinite dimensional representations
belonging to the class of representations with Dynkin labels $[0,\ldots,0,p]$.
We have shown that the superdimension of $[0,\ldots,0,p]$ coincides with the dimension of a $\mathfrak{so}(m-n)$ representation.
When the superdimension is finite, these representations could play a role in supergravity models.
Our technique is based on expansions of characters in terms of supersymmetric Schur functions.
In the process of studying these representations, we obtain new character expansions.
\end{abstract}

\maketitle

%%%%%%%%%%%%%%%%%%%%%%%%%%%%%%%%%%%%%%%%%%%%%%%%%%%%%%%%%%%%%%%%%%%%%%%%%%%%%%%
%%%%% section
%%%%%%%%%%%%%%%%%%%%%%%%%%%%%%%%%%%%%%%%%%%%%%%%%%%%%%%%%%%%%%%%%%%%%%%%%%%%%%%
\setcounter{equation}{0}
\section{Introduction} \label{sec:Introduction}%

Models of supergravity theory~\cite{Cremmer,GreenSchwarz} are often implicitly or explicitly based upon tensor representations of the orthosymplectic 
Lie superalgebra $\mathfrak{osp}(m|n)$~\cite{Baulieu,Preitschopf}.
Chiral spinors and self dual tensors of $\mathfrak{osp}(m|n)$ play an important role in such models.
These tensors are, however, infinite-dimensional. 
Nonetheless, the so-called superdimension of these tensors corresponds to the dimension of a finite-dimensional tensor 
of $\mathfrak{so}(m-n)$~\cite{STV}, thus paving the way for new covariant quantization schemes.

In~\cite{STV} we initiated the study of this correspondence between certain infinite-dimensional representations of $\mathfrak{osp}(m|n)$ and
finite-dimensional representations of $\mathfrak{so}(m-n)$.
Let us be more precise.
In terms of (the distinguished) Dynkin diagrams of $\mathfrak{osp}(m|n)$, the spinor representation has Dynkin labels 
$[0,0,\ldots,0,1]$ and the self dual tensor $[0,0,\ldots,0,2]$.
In~\cite{STV}, we treated the irreducible representations (irreps) with Dynkin labels $[0,0,\ldots,0,p]$, 
where $p$ is a positive integer (throughout this paper).

In the present paper, we shall first review some of the results of~\cite{STV}, and for this we need to recall some definitions and notations.
For all these developments, characters of a class of representations of $\mathfrak{osp}(m|n)$ play a prominent role.
Since the Lie superalgebras $\mathfrak{osp}(2m+1|2n)$ and $\mathfrak{osp}(2m|2n)$ both contain the general linear Lie superalgebra
$\mathfrak{gl}(m|n)$ as a subalgebra, it is convenient to express the characters of the infinite-dimensional $\mathfrak{osp}$-irreps as
an infinite sum of $\mathfrak{gl}(m|n)$ characters (given by supersymmetric Schur functions).
In~\cite{STV} this was done for the irreps $[0,0,\ldots,0,p]$ of $\mathfrak{osp}(2m(+1)|2n)$ (leading to a new character formula for the
case of $\mathfrak{osp}(2m|2n)$). 
In the current paper, we can extend this and obtain new character formulas for irreps of type $[0,\ldots,0,r,p-r]$ for $\mathfrak{osp}(2m|2n)$.

%%%%%%%%%%%%%%%%%%%%%%%%%%%%%%%%%%%%%%%%%%%%%%%%%%%%%%%%%%%%%%%%%%%%%%%%%%%%%%%
%%%%% section
%%%%%%%%%%%%%%%%%%%%%%%%%%%%%%%%%%%%%%%%%%%%%%%%%%%%%%%%%%%%%%%%%%%%%%%%%%%%%%%
\section{Definitions and notations}

\subsection{Partitions and (super)symmetric functions}

We need some basic notions on partitions and symmetric functions, see~\cite{Mac} as a standard reference.
A partition $\lambda=(\lambda_1,\lambda_2,\ldots,\lambda_n)$ of weight $|\lambda|$ and length $\ell(\lambda)\leq n$
is a sequence of non-negative integers satisfying the condition $\lambda_1\geq\lambda_2\geq\cdots\geq\lambda_n$, such that their
sum is $|\lambda|$, and $\lambda_i>0$ if and only if $i\leq \ell(\lambda)$. 
To each such partition there corresponds a Young diagram $F^\lambda$ consisting of $|\lambda|$ boxes arranged in $\ell(\lambda)$ left-adjusted
rows of lengths $\lambda_i$ for $i=1,2,\ldots,\ell(\lambda)$. 
For example, the Young diagram of $\lambda=(5,4,4,2)$ is given by
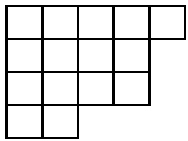
\[
\begin{Young}
&&&&\cr
&&&\cr
&&&\cr
&\cr
\end{Young}
\]
The conjugate partition $\lambda'$ corresponds to the Young diagram of $\lambda$ reflected about the main diagonal. 
For the above example, $\lambda'=(4,4,3,3,1)$.

If $\lambda,\mu$ are partitions, one writes $\lambda \supset \mu$ if the diagram of $\lambda$ contains that of $\mu$.
The difference $\lambda - \mu$ is called a skew diagram~\cite{Mac}. 
For example, if $\mu=(4,4,3)$, then the boxes of the skew diagram $\lambda - \mu$ are crossed in the following picture:
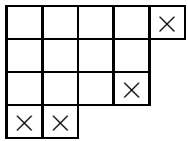
\[
\begin{Young}
&&&&$\times$\cr
&&&\cr
&&&$\times$\cr
$\times$&$\times$\cr
\end{Young}
\]
A skew diagram is a {\em horizontal strip} if it has at most one box in each column. 
The number of boxes of the horizontal strip is its length.
The above example is a horizontal strip of length~4.

Partitions are used to label symmetric functions. 
The Schur functions~\cite{Mac} or $S$-functions $s_\lambda(x)$ form a $\mathbb{Z}$-basis of the ring $\Lambda_n$ of symmetric polynomials with integer coefficients in the $n$ independent variables $x=(x_1,x_2,\ldots,x_n)$, where $\lambda$ runs over the set of all partitions of length at most $n$.
For a partition $\lambda$ with $\ell(\lambda)\leq n$, one has $s_\lambda(x) = \det (x_i^{\lambda_j+n-j})_{1\leq i,j\leq n}/\det (x_i^{n-j})_{1\leq i,j\leq n}$.
If $\ell(\lambda)>n$, one puts $s_\lambda(x)=0$. 

In terms of two sets of variables $x=(x_1,\ldots,x_m)$ and $y=(y_1,\ldots,y_n)$, one can define the ring $\Lambda_{m,n}$ of supersymmetric polynomials with integer coefficients~\cite{Berele}. This ring consists of all double symmetric polynomials in $x$ and $y$ (elements of $\Lambda_m\otimes \Lambda_n$) that satisfy the so-called cancellation property (i.e.\ when the substitution $x_1=t$, $y_1=-t$ is made in an element $p$ of $\Lambda_m\otimes \Lambda_n$, the resulting polynomial is independent of $t$). 
For a partition $\lambda$, one can define supersymmetric Schur polynomials $s_\lambda(x|y)$ belonging to $\Lambda_{m,n}$~\cite{Berele,King1983}.
These polynomials $s_\lambda(x|y)$ are zero when $\lambda_{m+1}>n$. 
Denote by ${\cal H}_{m,n}$ the set of all partitions with $\lambda_{m+1}\leq n$, i.e.\ the partitions (with their Young diagram) inside the $(m,n)$-hook. 
The set of $s_\lambda(x|y)$ with $\lambda\in{\cal H}_{m,n}$ forms a $\mathbb{Z}$-basis of $\Lambda_{m,n}$.

\subsection{Dimension, superdimension, $t$-dimension}

A finite-dimensional irreducible representation of the Lie algebra $\mathfrak{gl}(n)$ is characterized 
by a partition $\lambda$ with $\ell(\lambda)\leq n$. 
In terms of the standard basis $\epsilon_1,\ldots,\epsilon_n$ of the weight space of $\mathfrak{gl}(n)$, the highest weight of this
representation is $\sum_{i=1}^n \lambda_i \epsilon_i$, and the representation space will be denoted by $V_{\mathfrak{gl}(n)}^{\lambda}$. 
Weyl's character formula for such representations yields $\ch V_{\mathfrak{gl}(n)}^{\lambda}  = s_\lambda(x)$,
where $x_i=\hbox{e}^{\epsilon_i}$.

Just as the functions $s_\lambda(x)$ are characters of irreducible representations (or simple modules) of the Lie algebra $\mathfrak{gl}(n)$,
the supersymmetric Schur functions are characters of a class of simple modules of the Lie superalgebra $\mathfrak{gl}(m|n)$,
namely of the covariant representations~\cite{Berele}.
For a partition $\lambda\in{\cal H}_{m,n}$, the corresponding covariant representation will be denoted by $V_{\mathfrak{gl}(m|n)}^{\lambda}$.
In terms of the standard basis $\epsilon_1,\ldots,\epsilon_m,\ \delta_1,\ldots,\delta_n$ of the weight space of $\mathfrak{gl}(m|n)$, 
the highest weight of this representation is $\sum_{i=1}^m \lambda_i \epsilon_i + \sum_{j=1}^n \max(\lambda_j'-m,0)\delta_j$.
The main result of~\cite{Berele} is
\begin{equation}
\ch V_{\mathfrak{gl}(m|n)}^{\lambda} = s_\lambda(x|y),
\end{equation}
where $x_i=\hbox{e}^{\epsilon_i}$ and $y_j=\hbox{e}^{\delta_j}$.

Any Lie superalgebra $\mathfrak{g}$ is $\mathbb{Z}_2$-graded: $\mathfrak{g}=\mathfrak{g}_{\bar 0}\oplus\mathfrak{g}_{\bar 1}$.
A Lie superalgebra module or representation $V$ is also $\mathbb{Z}_2$-graded: $V=V_{\bar 0}\oplus V_{\bar 1}$. 
In our convention, the highest weight vector $v$ of $V$ will always be an even vector ($v\in V_{\bar 0}$).
When $V$ is finite-dimensional, one can speak of the dimension and superdimension of $V$:
\[
\dim V = \dim V_{\bar 0}+ \dim V_{\bar 1},\qquad \sdim V = \dim V_{\bar 0} - \dim V_{\bar 1}.
\]
Superdimension formulas for covariant representations of $\mathfrak{gl}(m|n)$ are known~\cite{King1983}. 
The result depends on whether $m$ is greater than, equal to, or less than $n$. 
It can be summarized as follows:
\begin{equation}
\sdim V_{\mathfrak{gl}(n+k|n)}^{\lambda} =\dim V_{\mathfrak{gl}(k)}^{\lambda}, \qquad
\sdim V_{\mathfrak{gl}(m|m+k)}^{\lambda} =(-1)^{|\lambda|}\dim V_{\mathfrak{gl}(k)}^{\lambda'}.
\label{dim-sdim}
\end{equation}
In particular, when $m=n$, $\sdim V_{\mathfrak{gl}(n|n)}^{\lambda} =0$ unless $\lambda$ is the zero partition $(0)$ (then 
$V_{\mathfrak{gl}(n|n)}^{(0)}$ is the trivial module with $\sdim V_{\mathfrak{gl}(n|n)}^{(0)} =1$).
Note that~\eqref{dim-sdim} implies: when $\ell(\lambda)>k$ then $\sdim V_{\mathfrak{gl}(n+k|n)}^{\lambda}=0$;
when $\lambda_1>k$ then $\sdim V_{\mathfrak{gl}(m|m+k)}^{\lambda}=0$.

Finally, let us introduce the notion of $t$-dimension of a Lie (super)algebra highest weight representation $V$.
This is nothing else but a specialization of the character of $V$, just like the $q$-dimension~\cite[Chapter~10]{Kac-book}.
Recall that the $q$-dimension of $V$, with highest weight $\Lambda$, is the specialization $F_1(\hbox{e}^{-\Lambda} \ch V)$, 
where $F_1$ is determined by
\[
F_1(\hbox{e}^{-\alpha_i})=q,
\]
and the $\alpha_i$'s are the simple roots of the Lie (super)algebra. 
So this corresponds to a gradation with respect to the simple roots.

The $t$-dimension is again a specialization $F(\hbox{e}^{-\Lambda} \ch V)$ of the character, but now $F$ is determined in a different way.
For a Lie algebra, of which the simple roots are commonly expressed in terms of the standard basis $\epsilon_1,\ldots,\epsilon_n$, 
one puts $F(\hbox{e}^{-\epsilon_i})=t$. 
For a Lie superalgebra, of which the simple roots are commonly expressed in terms of the standard basis $\epsilon_1,\ldots,\epsilon_m$,
$\delta_1,\ldots,\delta_n$, one puts $F(\hbox{e}^{-\epsilon_i})=t$ and $F(\hbox{e}^{-\delta_i})=t$ for the $t$-dimension, and
$F(\hbox{e}^{-\epsilon_i})=t$ and $F(\hbox{e}^{-\delta_i})=-t$ for the $t$-superdimension.

Let us clarify the meaning by means of an example.
Consider the orthogonal Lie algebra $\mathfrak{so}(2n+1)$, with simple roots $\epsilon_1-\epsilon_2, \ldots, \epsilon_{n-1}-\epsilon_n, \epsilon_n$,
and the representation $V$ with Dynkin labels $[0,\ldots,0,p]$, for which the highest weight is $(\frac{p}{2}, \ldots, \frac{p}{2})$ in
the $\epsilon$-basis. 
%This representation will be denoted by $V_{\mathfrak{so}(2n+1)}^{(p/2)^n}$.
For this representation, the character reads~\cite{parafermion, BG1}
\begin{equation}
\ch [0,\ldots,0,p]_{\mathfrak{so}(2n+1)} 
= (x_1\cdots x_n)^{-p/2} \sum_{\lambda_1\leq p,\; \ell(\lambda)\leq n} s_\lambda (x).
\label{e4}
\end{equation}
So the sum is over all partitions $\lambda$ such that the Young diagram of $\lambda$ fits inside the $n\times p$ rectangle, of width $p$ and height $n$.
Specializing this character according to $F$, one finds:
\begin{equation}
\dim_t [0,\ldots,0,p]_{\mathfrak{so}(2n+1)} 
%= \sum_{\lambda_1\leq p,\; \ell(\lambda)\leq n} s_\lambda(t,\ldots,t) 
= \sum_{\lambda_1\leq p,\; \ell(\lambda)\leq n} \dim V^\lambda_{\mathfrak{gl}(n)} t^{|\lambda|}.
\label{e5}
\end{equation}
When the character is expressed in terms of Schur functions, as in~\eqref{e4}, it yields in fact the branching of the representation 
according to $\mathfrak{so}(2n+1) \rightarrow \mathfrak{gl}(n)$.
When the character is specialized as in~\eqref{e5}, it is a polynomial in $t$ (or, in case of an infinite-dimensional representation, 
a formal power series in $t$) such that the coefficient of $t^k$ counts the dimension ``at level $k$'' according to the $\mathbb{Z}$-gradation
induced by the $\mathfrak{gl}(n)$ subalgebra of $\mathfrak{so}(2n+1)$.

\subsection{$t$-dimension for $\mathfrak{osp}(1|2n)$}

In this subsection we shall consider the $t$-dimension for a class of representations of $\mathfrak{g}=\mathfrak{osp}(1|2n)$.
Let us first fix some notation~\cite{Kac,Kac1,dict}. 
In the common basis $\delta_j$ for the weight space of $\mathfrak{osp}(1|2n)$, the odd roots are given by $\pm\delta_j$ ($j=1,\ldots,n$), and the even roots are $\delta_i-\delta_j$ ($i\ne j$) and $\pm(\delta_i+\delta_j)$. 
The simple roots are 
\begin{equation}
\delta_1-\delta_2,\  \delta_2-\delta_3, \ldots, \delta_{n-1}-\delta_n,\ \delta_n. 
\label{osp12n}
\end{equation}
The character specialization of the previous subsection corresponds to the 
following $\mathbb{Z}$-gradation of $\mathfrak{g}= \mathfrak{osp}(1|2n)$: 
$\mathfrak{g}=\mathfrak{g}_{-2}\oplus\mathfrak{g}_{-1}\oplus\mathfrak{g}_0\oplus\mathfrak{g}_{+1}\oplus\mathfrak{g}_{+2}$, where each
$\mathfrak{g}_j$ is spanned by the root vectors corresponding to the following roots:
\[
\begin{array}{ccccc}
\phantom{mm}\mathfrak{g}_{-2}\phantom{mm} &\phantom{mm}\mathfrak{g}_{-1}\phantom{mm} &\phantom{mm}\mathfrak{g}_{0}\phantom{mm} &\phantom{mm}\mathfrak{g}_{+1}\phantom{mm} &\phantom{mm}\mathfrak{g}_{+2}\phantom{mm} \\
-\delta_i-\delta_j & -\delta_i & \delta_i-\delta_j & \delta_i & \delta_i+\delta_j
\end{array}
\]
Note that $\mathfrak{g}_{0}=\mathfrak{gl}(n)$. 

We will consider infinite-dimensional highest weight representations $V$ of $\mathfrak{g}$, 
such that the action of $\mathfrak{g}_0=\mathfrak{gl}(n)$ on the highest weight vector $v$ of $V$ 
corresponds to a finite-dimensional $\mathfrak{g}_0$ module $V_0$.
Then the $\mathbb{Z}$-gradation of $\mathfrak{g}$ induces a $\mathbb{Z}$-gradation of $V$:
\[
V=V_0\oplus V_{-1} \oplus V_{-2} \oplus \cdots
\]
in terms of finite-dimensional $\mathfrak{g}_0$ modules, and the $t$-(super)dimension gives
\begin{equation}
\dim_t (V) = \sum_{i=0}^\infty \dim V_{-i}\ t^i, \quad \sdim_t (V) = \sum_{i=0}^\infty \dim V_{-i}\ (-t)^i=\dim_{-t} (V) .
\end{equation}

For reasons that will become clear, we will consider the irreducible highest weight representation with highest weight given by
$(-\frac{p}{2},-\frac{p}{2},\ldots, -\frac{p}{2})$ in the $\delta$-basis.
For this representation, the Dynkin labels are $[0,0,\ldots,0,-p]$. 
The structure and character of this representation have been determined in~\cite{paraboson}. 
Using the notation $x_i=\hbox{e}^{-\delta_i}$, one has:
\begin{equation}
\ch [0,0,\ldots,0,-p]_{\mathfrak{osp}(1|2n)} = (x_1\cdots x_n)^{p/2} \sum_{\lambda,\ \ell(\lambda)\leq p} s_\lambda(x).
\label{char-osp-1}
\end{equation}
This is an infinite sum over all partitions of length at most $p$. Since $s_\lambda(x)=0$ if $\ell(\lambda)>n$, the sum is actually over all partitions satisfying $\ell(\lambda)\leq\min(n,p)$.
Thus:
\begin{equation}
\dim_t [0,0,\ldots,0,-p]_{\mathfrak{osp}(1|2n)} = \sum_{\lambda,\ \ell(\lambda)\leq\min(n,p)} \dim V_{\mathfrak{gl}(n)}^{\lambda} t^{|\lambda|}.
\label{tdim-osp1}
\end{equation}
This infinite sum can be rewritten in an alternative form, see~\cite{STV}.

%%%%%%%%%%%%%%%%%%%%%%%%%%%%%%%%%%%%%%%%%%%%%%%%%%%%%%%%%%%%%%%%%%%%%%%%%%%%%%%
%%%%% section
%%%%%%%%%%%%%%%%%%%%%%%%%%%%%%%%%%%%%%%%%%%%%%%%%%%%%%%%%%%%%%%%%%%%%%%%%%%%%%%
\section{Superdimensions for $\mathfrak{osp}(2m+1|2n)$} 
\label{sec3}

Consider the Lie superalgebra $B(m,n)=\mathfrak{osp}(2m+1|2n)$, with the distinguished set
of simple roots in the $\epsilon$-$\delta$-basis~\cite{Kac,dict} 
\begin{equation}
\delta_1-\delta_2,\  \ldots, \delta_{n-1}-\delta_n,\ \delta_n-\epsilon_1,\ \epsilon_1-\epsilon_2,\ \ldots, \epsilon_{m-1}-\epsilon_m,\ \epsilon_m. 
\label{osp2m12n}
\end{equation}
Also in this case there exists a useful $\mathbb{Z}$-gradation of $\mathfrak{g}= \mathfrak{osp}(2m+1|2n)$: 
$\mathfrak{g}=\mathfrak{g}_{-2}\oplus\mathfrak{g}_{-1}\oplus\mathfrak{g}_0\oplus\mathfrak{g}_{+1}\oplus\mathfrak{g}_{+2}$, where each
$\mathfrak{g}_j$ is spanned by the root vectors corresponding to the following roots:
\[
\begin{array}{ccccc}
\phantom{mm}\mathfrak{g}_{-2}\phantom{mm} &\phantom{mm}\mathfrak{g}_{-1}\phantom{mm} &\phantom{mm}\mathfrak{g}_{0}\phantom{mm} &\phantom{mm}\mathfrak{g}_{+1}\phantom{mm} &\phantom{mm}\mathfrak{g}_{+2}\phantom{mm} \\
-\delta_i-\delta_j & -\delta_i & \delta_i-\delta_j & \delta_i & \delta_i+\delta_j\\
-\epsilon_i-\epsilon_j\ (i\ne j) & -\epsilon_i & \epsilon_i-\epsilon_j & \epsilon_i & \epsilon_i+\epsilon_j\ (i\ne j)\\
-\epsilon_i-\delta_j & & \pm(\epsilon_i-\delta_j) & & \epsilon_i+\delta_j
\end{array}
\]
So $\mathfrak{g}_{0}=\mathfrak{gl}(m|n)$, and this gradation corresponds to the $t$-(super)dimension introduced earlier.

Let us consider the irreducible highest weight representation with highest weight given by
$(\frac{p}{2},\ldots,\frac{p}{2};-\frac{p}{2},\ldots, -\frac{p}{2})$ in the $\epsilon$-$\delta$-basis.
This representation has Dynkin labels $[0,0,\ldots,0,p]$.
Using $x_i=\hbox{e}^{-\epsilon_i}$, $y_i=\hbox{e}^{-\delta_i}$, the following character formula holds~\cite{parast,STV}:
\begin{equation}
\ch [0,\ldots,0,p]_{\mathfrak{osp}(2m+1|2n)} = (y_1\cdots y_n/x_1\cdots x_m)^{p/2} \sum_{\lambda,\ \lambda_1\leq p} s_\lambda(x|y).
\label{char-Bmn}
\end{equation}
So here the sum is over all partitions $\lambda$ inside the $(m,n)$-hook (otherwise $s_\lambda(x|y)$ is zero anyway) with 
$\lambda_1\leq p$, or equivalently $\ell(\lambda')\leq p$.

In order to determine $\sdim_t [0,\ldots,0,p]_{\mathfrak{osp}(2m+1|2n)}$, one should (apart from the factor in front of the above
sum) specify $x_i=t$ and $y_j=-t$ in the above character, and so one finds
\begin{align}
\sdim_t [0,\ldots,0,p]_{\mathfrak{osp}(2m+1|2n)} &= \sum_{\lambda,\ \lambda_1\leq p} s_\lambda(t,\ldots,t|-t,\ldots,-t) \nonumber\\
&= \sum_{\lambda,\ \lambda_1\leq p} s_\lambda(1,\ldots,1|-1,\ldots,-1)\, t^{|\lambda|}  \nonumber\\
&= \sum_{\lambda,\ \lambda_1\leq p} \sdim V_{\mathfrak{gl}(m|n)}^{\lambda} \, t^{|\lambda|}. 
\label{sdim-Bmn}
\end{align}
Using the properties of $\mathfrak{gl}(m|n)$ superdimensions, this leads to the following three cases.

\vskip 1mm
\noindent {\bf Case 1: $m=n$, $\mathfrak{osp}(2n+1|2n)$.} 
All superdimensions of covariant representations of $\mathfrak{gl}(n|n)$ are zero, except when $\lambda=(0)$. Hence:
\begin{equation}
\sdim_t [0,\ldots,0,p]_{\mathfrak{osp}(2n+1|2n)} = 1.
\end{equation}

\vskip 1mm
\noindent {\bf Case 2: $m=n+k$, $\mathfrak{osp}(2n+2k+1|2n)$.}
Now it follows directly from~\eqref{sdim-Bmn} and~\eqref{dim-sdim} that
\begin{equation}
\sdim_t [0,\ldots,0,p]_{\mathfrak{osp}(2m+1|2n)} = \sum_{\lambda,\ \lambda_1\leq p} \dim V_{\mathfrak{gl}(k)}^{\lambda} \, t^{|\lambda|}
= \sum_{\lambda,\ \lambda_1\leq p,\ \ell(\lambda)\leq k} \dim V_{\mathfrak{gl}(k)}^{\lambda} \, t^{|\lambda|}.
\label{sdim-Bk1}
\end{equation}
This coincides with expression~\eqref{e5}. Hence we can write
\begin{equation}
\sdim_t [0,0,\ldots,0,p]_{\mathfrak{osp}(2n+2k+1|2n)} = \dim_t [0,\ldots,0,p]_{\mathfrak{so}(2k+1)}.
\label{sdim-Bk3}
\end{equation}

\vskip 1mm
\noindent {\bf Case 3: $n=m+k$, $\mathfrak{osp}(2m+1|2m+2k)$.} One finds:
\begin{equation}
\sdim_t [0,\ldots,0,p]_{\mathfrak{osp}(2m+1|2n)} = \sum_{\lambda,\ \lambda_1\leq p,\ \lambda_1\leq k}
(-1)^{|\lambda|} \dim V_{\mathfrak{gl}(k)}^{\lambda'} \, t^{|\lambda|}
= \sum_{\mu,\ \ell(\mu)\leq\min(p,k)} \dim V_{\mathfrak{gl}(k)}^{\mu} \, (-t)^{|\mu|}.
\label{sdim-sBk1}
\end{equation}
The right hand side is the same expression as~\eqref{tdim-osp1}, so 
\begin{equation}
\sdim_t [0,0,\ldots,0,p]_{\mathfrak{osp}(2m+1|2m+2k)} = \dim_{-t} [0,\ldots,0,-p]_{\mathfrak{osp}(1|2k)}.
\label{sdim-sBk3}
\end{equation}

So in all three cases, the superdimension for $\mathfrak{osp}(2m+1|2n)$ simplifies and reduces 
to a dimension of $\mathfrak{so}(2m+1-2n)$ or $\mathfrak{osp}(1|2n-2m)$.

%%%%%%%%%%%%%%%%%%%%%%%%%%%%%%%%%%%%%%%%%%%%%%%%%%%%%%%%%%%%%%%%%%%%%%%%%%%%%%%
%%%%% section
%%%%%%%%%%%%%%%%%%%%%%%%%%%%%%%%%%%%%%%%%%%%%%%%%%%%%%%%%%%%%%%%%%%%%%%%%%%%%%%
\section{Superdimensions for $\mathfrak{osp}(2m|2n)$ and new characters} 

For $D(m,n)=\mathfrak{osp}(2m|2n)$, the distinguished set of simple roots in the $\epsilon$-$\delta$-basis is
\begin{equation}
\delta_1-\delta_2,\  \ldots, \delta_{n-1}-\delta_n,\ \delta_n-\epsilon_1,\ \epsilon_1-\epsilon_2,\ \ldots, \epsilon_{m-2}-\epsilon_{m-1},\ \epsilon_{m-1}-\epsilon_m,\ \epsilon_{m-1}+\epsilon_m. 
\label{osp2m2n}
\end{equation}
It will be helpful to see $D(m,n)$ as a subalgebra of $B(m,n)$. 
In fact, using the $\mathbb{Z}$-gradation 
$\mathfrak{g}_{-2}\oplus\mathfrak{g}_{-1}\oplus\mathfrak{g}_0\oplus\mathfrak{g}_{+1}\oplus\mathfrak{g}_{+2}$ 
of $\mathfrak{g}=\mathfrak{osp}(2m+1|2n)$ introduced in the previous section, 
it is easy to see that $\mathfrak{osp}(2m|2n) = \mathfrak{g}_{-2}\oplus\mathfrak{g}_0\oplus\mathfrak{g}_{+2}$, with root structure
as in Section~\ref{sec3}.

For the irreducible highest weight representation of $\mathfrak{osp}(2m|2n)$ with highest weight given by
$(\frac{p}{2},\ldots,\frac{p}{2};-\frac{p}{2},\ldots, -\frac{p}{2})$, 
with Dynkin labels are $[0,0,\ldots,0,p]$, the character was determined in~\cite{STV}:
\begin{equation}
\ch [0,\ldots,0,p]_{\mathfrak{osp}(2m|2n)} = (y_1\cdots y_n/x_1\cdots x_m)^{p/2} 
\sum_{\lambda\in{\cal B},\ \lambda_1\leq p} s_\lambda(x|y).
\label{char-Dmn}
\end{equation}
Herein, ${\cal B}$ denotes the set of partitions for which each part appears twice (including the zero partition).
Thus, one finds
\begin{equation}
\sdim_t [0,\ldots,0,p]_{\mathfrak{osp}(2m|2n)} = 
\sum_{\lambda\in{\cal B},\ \lambda_1\leq p} \sdim V_{\mathfrak{gl}(m|n)}^{\lambda} \, t^{|\lambda|}. 
\label{sdim-Dmn}
\end{equation}
This expression allows once again to deduce superdimension formulas in three cases: $m=n$, $m>n$ and $m<n$, see~\cite{STV}.
Let us give here the formula for $m>n$, i.e.\ $m=n+k$, or $\mathfrak{osp}(2n+2k|2n)$.
From~\eqref{sdim-Dmn} one has:
\begin{equation}
\sdim_t [0,\ldots,0,p]_{\mathfrak{osp}(2m|2n)} = \sum_{\lambda\in{\cal B},\ \lambda_1\leq p} \dim V_{\mathfrak{gl}(k)}^{\lambda} \, t^{|\lambda|}
= \sum_{\lambda\in{\cal B},\ \lambda_1\leq p,\ \ell(\lambda)\leq k} \dim V_{\mathfrak{gl}(k)}^{\lambda} \, t^{|\lambda|}.
\label{sdim-Dk1}
\end{equation}
And thus, using known characters of $\mathfrak{so}(2k)$~\cite{STV}:
\begin{equation}
\sdim_t [0,0,\ldots,0,p]_{\mathfrak{osp}(2n+2k|2n)} = \left\{
\begin{array}{ll}
\displaystyle \dim_t [0,\ldots,0,0,p]_{\mathfrak{so}(2k)} & \hbox{ for }k\hbox{ even}. \\
\displaystyle \dim_t [0,\ldots,0,p,0]_{\mathfrak{so}(2k)} & \hbox{ for }k\hbox{ odd}. 
\end{array} \right.
\label{sdim-Dk3}
\end{equation}
Here, the convention for the order of the simple roots of $\mathfrak{so}(2k)$ is $\epsilon_1-\epsilon_2,
\ldots, \epsilon_{k-1}-\epsilon_k,\epsilon_{k-1}+\epsilon_k$.

At this point, we can make some interesting observations and additions to the results obtained in~\cite{STV}.
For this, let us first consider the representations appearing here for $\mathfrak{so}(2k+1)$ and $\mathfrak{so}(2k)$.
In~\eqref{e4} we obtained
\begin{equation}
\ch [0,\ldots,0,p]_{\mathfrak{so}(2k+1)} = (x_1\cdots x_k)^{-p/2} \sum_{\lambda_1\leq p,\; \ell(\lambda)\leq k} s_\lambda (x).
\end{equation}
Essentially, this is the branching $\mathfrak{so}(2k+1) \supset \mathfrak{gl}(k)$. 
But for this inclusion, there is an intermediate subalgebra: $\mathfrak{so}(2k+1) \supset \mathfrak{so}(2k) \supset \mathfrak{gl}(k)$. 
From Weyl's character formula, it is easy to deduce the branching of the above $\mathfrak{so}(2k+1)$ representation with respect to $\mathfrak{so}(2k)$:
\begin{equation}
\ch [0,\ldots,0,p]_{\mathfrak{so}(2k+1)} = \sum_{r=0}^p \ch [0,\ldots,r,p-r]_{\mathfrak{so}(2k)}.
\label{e28}
\end{equation}
The $\mathfrak{so}(2k)$ representations that appeared earlier, with expressions in terms of Schur functions, were $[0,\ldots,0,p]$ and $[0,\ldots,0,p,0]$.
So the question is now: how to write the character of the other $\mathfrak{so}(2k)$ representations $[0,\ldots,r,p-r]$ as a sum of Schur functions?
Or in other words, what is the branching $\mathfrak{so}(2k) \supset \mathfrak{gl}(k)$ for these representations?
The answer is:

\noindent {\bf Theorem.} {\em For $k$ even, one has
\begin{equation}
\ch [0,\ldots,0,r,p-r]_{\mathfrak{so}(2k)} = (x_1\cdots x_k)^{-p/2} 
\sum_{\lambda_1\leq p,\; \ell(\lambda)\leq k;\; \lambda\in{\cal B}_r} s_\lambda (x).
\end{equation}
Herein, ${\cal B}_r$ stands for the set of partitions of ${\cal B}$ to which a horizontal strip of length~$r$ is attached.
(Recall that ${\cal B}$ is the set of partitions for which each part appears twice.)
The first condition ($\lambda_1\leq p,\; \ell(\lambda)\leq k$) means that (the Young diagram of) $\lambda$ fits inside the $k\times p$ rectangle.
Similarly, for $k$ odd:
\begin{equation}
\ch [0,\ldots,0,r,p-r]_{\mathfrak{so}(2k)} = (x_1\cdots x_k)^{-p/2} 
\sum_{\lambda_1\leq p,\; \ell(\lambda)\leq k;\; \lambda\in{\cal B}_{p-r}} s_\lambda (x).
\end{equation}
}

We have not found the above result in the literature. 
The actual proof is rather technical. It can be obtained using the branching rules for $\mathfrak{so}(2k) \supset \mathfrak{gl}(k)$ described in~\cite{KW}.
Note that, in accordance with~\eqref{e28}, the union of all partitions of ${\cal B}_r$ in the $k\times p$ rectangle, for $r=0,1,\ldots,p$, 
is equal to the set of all partitions in the rectangle.

But now we can extend the analogy that we observed between representations $[0,\ldots,0,p]$ of $\mathfrak{osp}(2m|2n)$ 
and those of $\mathfrak{so}(2k)$ for $m=n+k$. 
This leads to the following

\noindent {\bf Conjecture.} {\em For $|m-n|$ even, one has
\begin{equation}
\ch [0,\ldots,0,r,p-r]_{\mathfrak{osp}(2m|2n)} = (y_1\cdots y_n/x_1\cdots x_m)^{p/2} 
\sum_{\lambda_1\leq p,\; \lambda\in{\cal B}_r} s_\lambda (x/y).
\end{equation}
So in this case we have an expansion as an infinite sum of supersymmetric Schur functions,
labeled by partitions $\lambda$ inside the $(m,n)$-hook, of width at most $p$, and belonging to ${\cal B}_r$.}

For $|m-n|$ odd, the result is similar, with ${\cal B}_r$ replaced by ${\cal B}_{p-r}$.

To conclude the paper, we have analyzed characters and superdimensions for representations of the form $[0,\ldots,0,p]$ for $\mathfrak{osp}(2m+1|2n)$, and of the form $[0,\ldots,0,r,p-r]$ for $\mathfrak{osp}(2m|2n)$.
It should be noted that characters for more general $\mathfrak{osp}(m|n)$ tensors have been studied in~\cite{Cheng}.
However, the formulas in~\cite{Cheng} lead to alternating series of $S$-functions, which are not as easy to handle as the characters obtained here.

%%%%%%%%%%%%%%%%%%%%%%%%%%%%%%%%%%%%%%%%%%%%%%%%%%%%%%%%%%%%%%%%%%%%%%%%%%%%%%%
%%%%% references
%%%%%%%%%%%%%%%%%%%%%%%%%%%%%%%%%%%%%%%%%%%%%%%%%%%%%%%%%%%%%%%%%%%%%%%%%%%%%%%


\begin{thebibliography}{99}

\bibitem{Cremmer} 
E.\ Cremmer, B.\ Julia and J.\ Scherk, 
%Supergravity Theory in Eleven-Dimensions
{Phys.\ Lett.} {\bf 76B}, 409 (1978).

\bibitem{GreenSchwarz}
M.B.\ Green and J.H.\ Schwarz, 1983
%Extended Supergravity in 10 Dimensions
{Phys.\ Lett.\ B} {\bf 122}, 143 (1983).

\bibitem{Baulieu}
L.\ Baulieu and J.\ Thierry-Mieg,
%Covariant quantization of non-Abelian antisymmetric tensor gauge theories
{Nucl.\ Phys.\ B} {\bf 228}, 259 (1983).

\bibitem{Preitschopf}
C.R.\ Preitschopf, T.\ Hurth, P.\ van Nieuwenhuizen and A.\ Waldron,
%Title = {{Osp(1/8)-gravity}},
{Nucl.\ Phys.\ B} {\bf 56B}, 310 (1997).

\bibitem{STV}
N.I.\ Stoilova, J.\ Thierry-Mieg and J.\ Van der Jeugt,
% Extension of the osp(m|n)~so(m-n) correspondence to the infinite-dimensional chiral spinors and self dual tensors,
{J.\ Phys.\ A} {\bf 50}, 155201 (2017).

\bibitem{Mac}
I.G.\ Macdonald,
{\em Symmetric Functions and Hall Polynomials. 2nd edition} (Oxford University Press, Oxford, 1995).

\bibitem{Berele}
A.\ Berele and A.\ Regev, 
%Hook Young diagrams with applications to combinatorics and to representations of Lie superalgebras
{\em Adv.\ Math.} {\bf 64}, 118 (1987). 

\bibitem{King1983}
R.C.\ King,
%Supersymmetric functions and the Lie supergroup $\rm {U(m/n)}$
{\em Ars.\ Combin.} {\bf 16A}, 269 (1983).

\bibitem{Kac-book}
V.G.\ Kac,
{\em Infinite dimensional Lie algebras. 2nd edition} (Cambridge University Press, Cambridge, 1985).

\bibitem{parafermion}
N.I.\ Stoilova and J.\ Van der Jeugt, 
%The parafermion Fock space and explicit $\mathfrak{so}(2n+1)$ representations
{J.\ Phys A} {\bf 41}, 075202 (2008).

\bibitem{BG1}
A.J.\ Bracken and H.S.\ Green,
%Algebraic Identities for Parafermi Statistics of Given Order
{Nuovo Cim.} {\bf 9}, 349 (1972).

\bibitem{Kac}
V.G.\ Kac, 
%Lie superalgebras 
{Adv.\ Math.} {\bf 26}, 8 (1977).

\bibitem{Kac1}
V.G.\ Kac,
%Representations of Classical Lie Superalgebras 
{Lect.\ Notes in Math.} {\bf 626}, 597 (1978). 

\bibitem{dict}
L.\ Frappat, A.\ Sciarrino and P.\ Sorba,
{\em Dictionary on Lie Algebras and Superalgebras} (Academic Press, London, 2000).

\bibitem{paraboson}
S.\ Lievens, N.I.\ Stoilova and J.\ Van der Jeugt,
%The paraboson Fock space and unitary irreducible representations of the Lie superalgebra $\mathfrak{osp}(1|2n)$
{Commun.\ Math.\ Phys.} {\bf 281}, 805 (2008).

\bibitem{parast}
N.I.\ Stoilova and J.\ Van der Jeugt,   
%A class of infinite-dimensional representations of the Lie superalgebra $\mathfrak{osp}(2m+1|2n)$ and the parastatistics Fock space
{J.\ Phys A} {\bf 48}, 155202  (2015).

\bibitem{KW}
R.C.\ King and B.G.\ Wybourne,
{J.\ Math.\ Phys.}  {\bf 41}, 5002 (2000).

\bibitem{Cheng}
S.J.\ Cheng and R.B.\ Zhang,
%Howe duality and combinatorial character formula for orthosymplectic Lie superalgebras
{Adv.\ Math.} {\bf 182}, 124 (2004).


\end{thebibliography}
\end{document}